\documentclass[11pt,twoside]{article}
\usepackage{asp2010}

\resetcounters

\begin{document}

\title{Multi-Wavelength Study of the 2008-2009 Outburst of V1647 Ori}
\author{D. Garc\'{\i}a-Alvarez$^{1,2}$, N.J. Wright$^3$, J.J. Drake$^3$, P. Abraham$^4$, B.G. Anandarao$^5$, V. Kashyap$^3$, A. Kospal$^6$, M. Kun$^4$, M. Marengo$^7$, A. Moor$^4$, S. Peneva$^8$, E. Semkov$^8$, V. Venkat$^5$, and J. Sanz-Forcada$^9$
\affil{$^1$IAC, Tenerife, Spain}
\affil{$^2$GRANTECAN, La Palma, Spain}
\affil{$^3$Harvard-Smithsonian CfA, MA, USA}
\affil{$^4$Konkoly Observarory, Budapest, Hungary}
\affil{$^5$Astronomy and Astrophysics Division, PLR, Ahmedabad, India}
\affil{$^6$Leiden University, Leiden, The Netherlands}
\affil{$^7$Department of Physics and Astronomy, Iowa State University, IA, USA}
\affil{$^8$Institute of Astronomy and National Astronomical Observatory Bulgarian Academy of Sciences, Sofia, Bulgaria}
\affil{$^9$Centro de Astrobiología, CSIC-INTA,Spain}}

\begin{abstract}
V1647 Ori is a young eruptive variable star, illuminating a reflection nebula (McNeil's Nebula). It underwent an outburst in 2003 before fading back to its pre-outburst brightness in 2006. In 2008, V1647 Ori underwent a new outburst. The observed properties of the 2003-2006 event are different in several respects from both the EXor and FUor type outbursts, and suggest that this star might represent a new class of eruptive young stars, younger and more deeply embedded than EXors, and exhibiting variations on shorter time scales than FUors. In outburst, the star lights up the otherwise invisible McNeil's nebular - a conical cloud likely accumulated from previous outbursts. We present follow-up photometric as well as optical and near-IR spectroscopy of the nebula obtainted during the 2008-2009 outburst. We will also present results from contemporaneous X-ray observations. These multi-wavelength observations of V1647 Ori, obtained at this key early stage of the outburst, provide a snapshot of the "lighting up" of the nebula, probe its evolution through the event, and enable comparison with the 2003-2006 outburst.
\end{abstract}

\section{Introduction}
Eruptive young stellar objects form a small but spectacular class of pre-main-sequence stars. Traditionally, they are divided into two groups. FU Orionis-type stars (FUors). Their spectral type is F-G giant according to the optical spectrum, and K-M giant/supergiant at near-infrared (NIR) wavelengths. The second group, called EX Lupi-type stars (EXors), belongs to the T Tauri class.  Their spectral type is K or M dwarf. Their outbursts are thought to be the consequence of a sudden and steep increase of the mass accretion rate onto the central star, which changes from those commonly found around T Tauri stars ($~10^{-7}$ $M_{\odot} yr^{-1}$) into values upto 10$^{3}$ times larger. Statistical studies suggest that young low-mass stars experience several FU Orionis like outbursts during the early phase of stellar evolution. The emergence of a new pre-main-sequence outburst object is then a unique opportunity to address the physical process that occurs in the disk's interior. For this reason several astronomers have focused their attention on the recent outbursts of V1647 Ori.

V1647 Ori is a young eruptive star known to be the illuminating source of McNeil's Nebula, a reflecting nebula \citep{mcneil04}. In the months following the discovery, the star, located in the dark cloud Lynds 1630 - a region of active star formation in the Orion B complex, showed an increase of its optical/IR brightness of up to 6 magnitudes. The outburst has been observed from the X-ray regime \citep[e.g.][]{grosso05} to infrared wavelength \citep[e.g.][]{muzerolle05}. Four months after the onset of the outburst, the brightness rise stopped and the magnitude remained relatively constant. Two years after the onset it started a fast fading phase in the optical light of V1647 Ori. The system is further characterized by a red energy distribution and by many emission lines in its optical and near-IR spectrum. Most of the Hydrogen lines exhibit a P-Cygni profile, which indicates mass outflow in a wind. Its optical and near-IR spectrum does not resemble any other previous spectra of FUors or EXors objects. In August 2008, V1647 Ori, onset another outburst just four years before the previous 2004 outburst. The 2-3 years duration of the outburst, its recurrence on a timescale of decades and the "peculiar" spectrum of V1647 Ori, are important clues for the comprehension of outburst events in pre-main-sequence stars. Table 1 compare the main features of FUors, EXors and V1647 Ori.

\begin{table}[!ht]
\caption{Typical value of outburst from pre-main-sequence stars \citep{fedele07}.}
\smallskip
\begin{center}
{\tiny
\begin{tabular}{llll}
\tableline
\noalign{\smallskip}
 & FUors & EXors & V1647 Ori \\
\noalign{\smallskip}
\tableline
\noalign{\smallskip}
Outburst duration [yr] & $>$10 & $~$1 & 2.6 \\
Outburst recurrence [yr] & $>$200  & 5-10 & 37 \\
Mass accreted during an outburst [M$_{\odot}$] & $>$10$^{-3}$  &10$^{-6}$-10$^{-5}$  & 2.5x10$^{-5}$ \\
Magnitude variation [optical mag] & 4-6 & 2-5 & $~$6 \\
Accretion Luminosity [L$_{\odot}$] & few 10$^{2}$ & $>$25 & 44 \\
Outburst accretion rate [M$_{\odot}$ yr$^{-1}$] & 10$^{-4}$ &10$^{-6}$-10$^{-5}$  & 10$^{-5}$ \\
Envelope infall rate [M$_{\odot}$ yr$^{-1}$] & 5x10$^{-6}$ & 10$^{-7}$-10$^{-6}$ & 7x10$^{-7}$ \\
Wind velocity [Km s$^{-1}$] & $>$300 & 200-400 & 300-400 \\
Mass loss rate [M$_{\odot}$ yr$^{-1}$] &10$^{-6}$-10$^{-5}$  & 10$^{-8}$-10$^{-6}$ & 4x10$^{-8}$ \\
Spectral features & absorption spectrum & emission line spectrum & emission line spectrum, \\
 &F/G-type supergiant like& T Tauri like, H$\alpha $ inverse P Cyg& 	strong H$\alpha $ emission (P Cygni)\\
  & like deep CO absorption& CO abs./em., Br$\gamma$ emission & CO abs./em., Br$\gamma$ emission \\
  & & & 	Forbidden lines in fading phase \\
Notes &  &  & X-ray variability \\
\noalign{\smallskip}
\tableline
\end{tabular}
}
\end{center}
\end{table}
\begin{figure}[]
\begin{center}
\includegraphics[scale=0.4, angle=90]{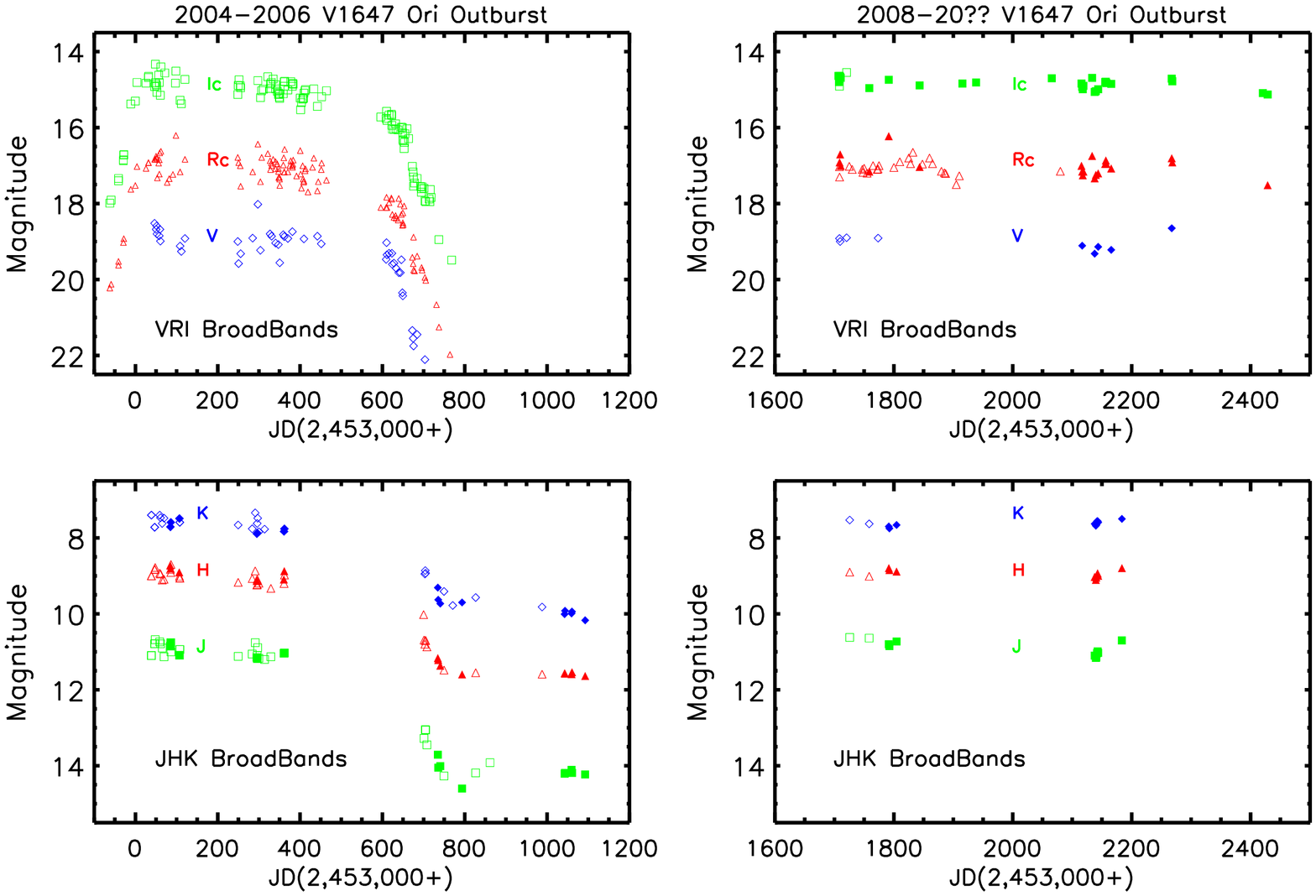}
\end{center}
\caption{Optical (VRI) and NIR (JHK), light curve of V1647 Ori during the 2004 and 2008 outburst. Data are from: filled symbols this work; open symbols from \citet{semkov06}; \citet{kospal05}; \citet{acosta-pulido07}; \citet{briceno04}; \citet{mcgehee04}; \citet{ojha06}; \citet{fedele07}; \citet{reipurth04}; \citet{chochol06}; \citet{aspin08}; \citet{ojha08}; \citet{kaurav10}; \citet{kun08}. An offset was applied to some of the imported data in order to reach a similar magnitude level with our data.}
\end{figure}
\begin{figure}[]
\begin{center}
\includegraphics[scale=0.4, angle=90]{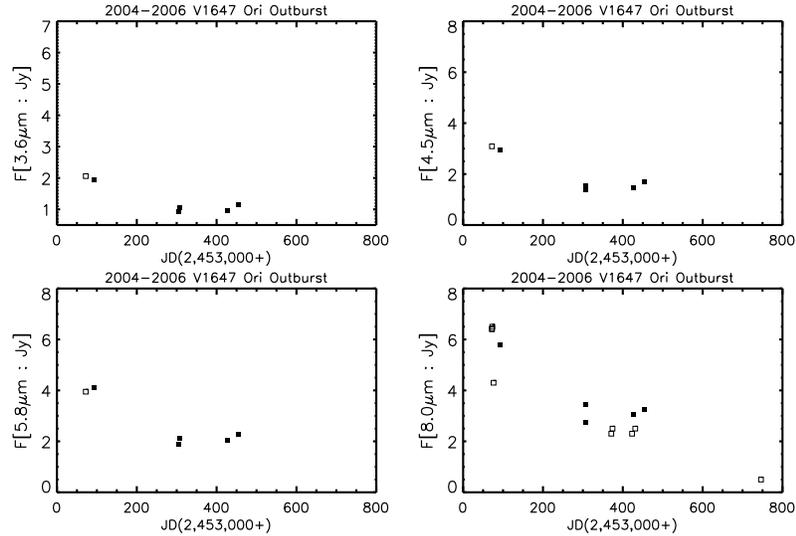}
\end{center}
\caption{3.6, 4.5, 5.8 and 8.0microns lightcurve of V1647 Ori of the 2004 outburst. Data are from: filled symbols this work; open symbols from \citet{muzerolle05}; \citet{fedele07}.}
\end{figure}
\begin{figure}[]
\begin{center}
\includegraphics[scale=0.4, angle=90]{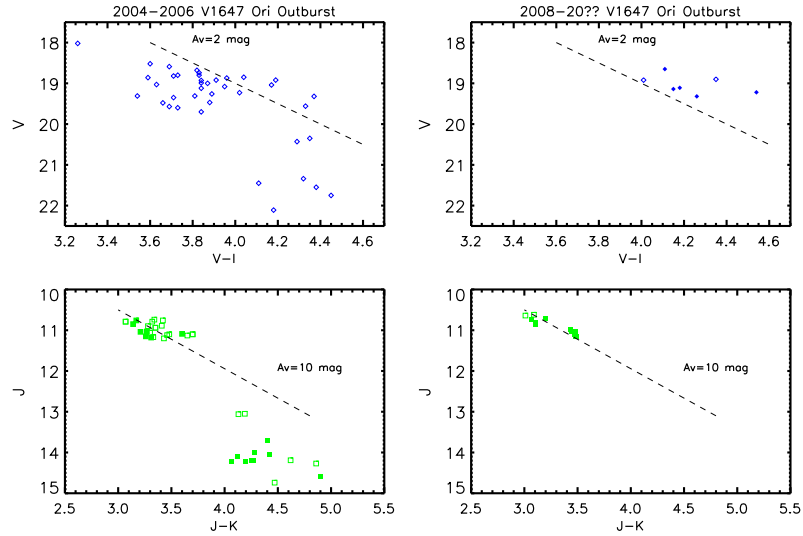}
\end{center}
\caption{Relationship between the color indices and magnitudes of V1647 Ori during the 2004 and 2008 outburst. Top, V vs. V-I; Bottom, J vs. J-K. Data are from: filled symbols this work; open symbols same as previous plot. The dotted lines indicate the slope of the normal interstellar reddening \citep{cohen81}.}
\end{figure}
\begin{figure}[!ht]
\begin{center}
\includegraphics[scale=0.4, angle=90]{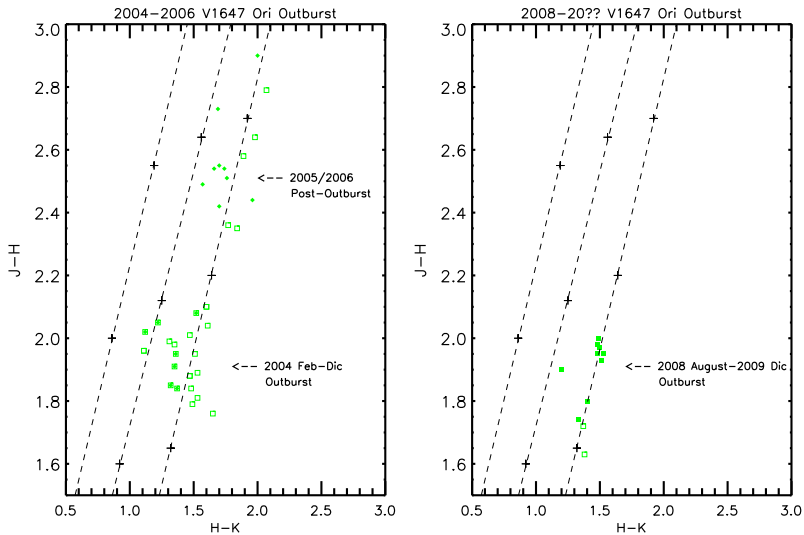}
\end{center}
\caption{J-H vs H-K diagram showing the location of V1647 Ori as a function on time related to the 2004 and 2008 outburst. The dashed straight lines represent the reddening vectors \citep{rieke85}. The crosses on the dashed lines are separated by AV=5 mag. Note that a general movement along a reddening vector can be seen during the outburst. The source had moved towards the tip of the locus of classical T Tauri stars \citep{meyer97} during 2004 and 2008. This suggests that V1647 Ori has NIR colour similar to a dereddened T Tauri star. The J-H and H-K colours indicate that the circumstellar matter of AV$\sim$8 mag was probably cleared in the eruptions. }
\end{figure}
\begin{figure}[]
\begin{center}
\includegraphics[scale=0.35]{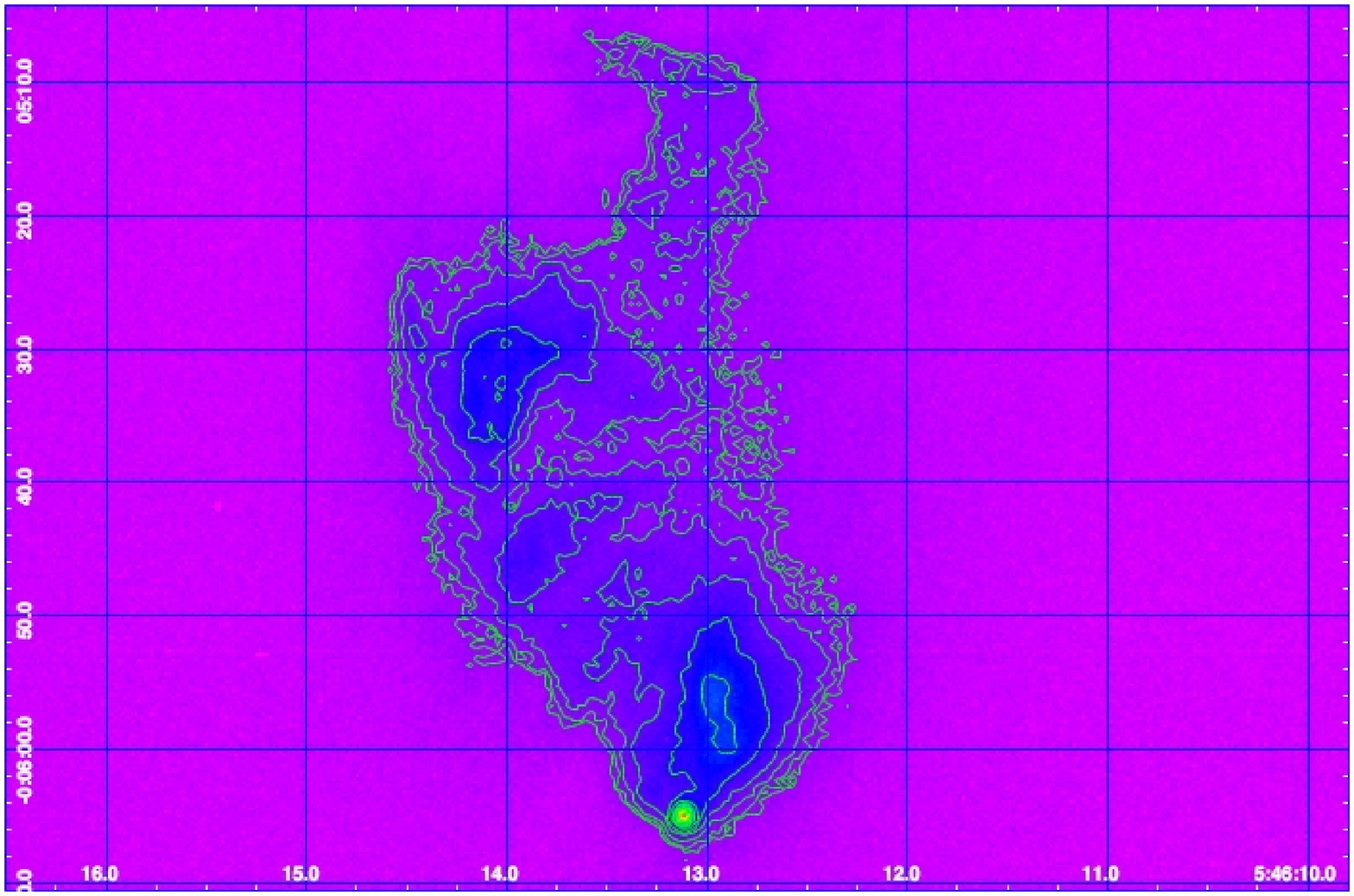}
\includegraphics[scale=0.3]{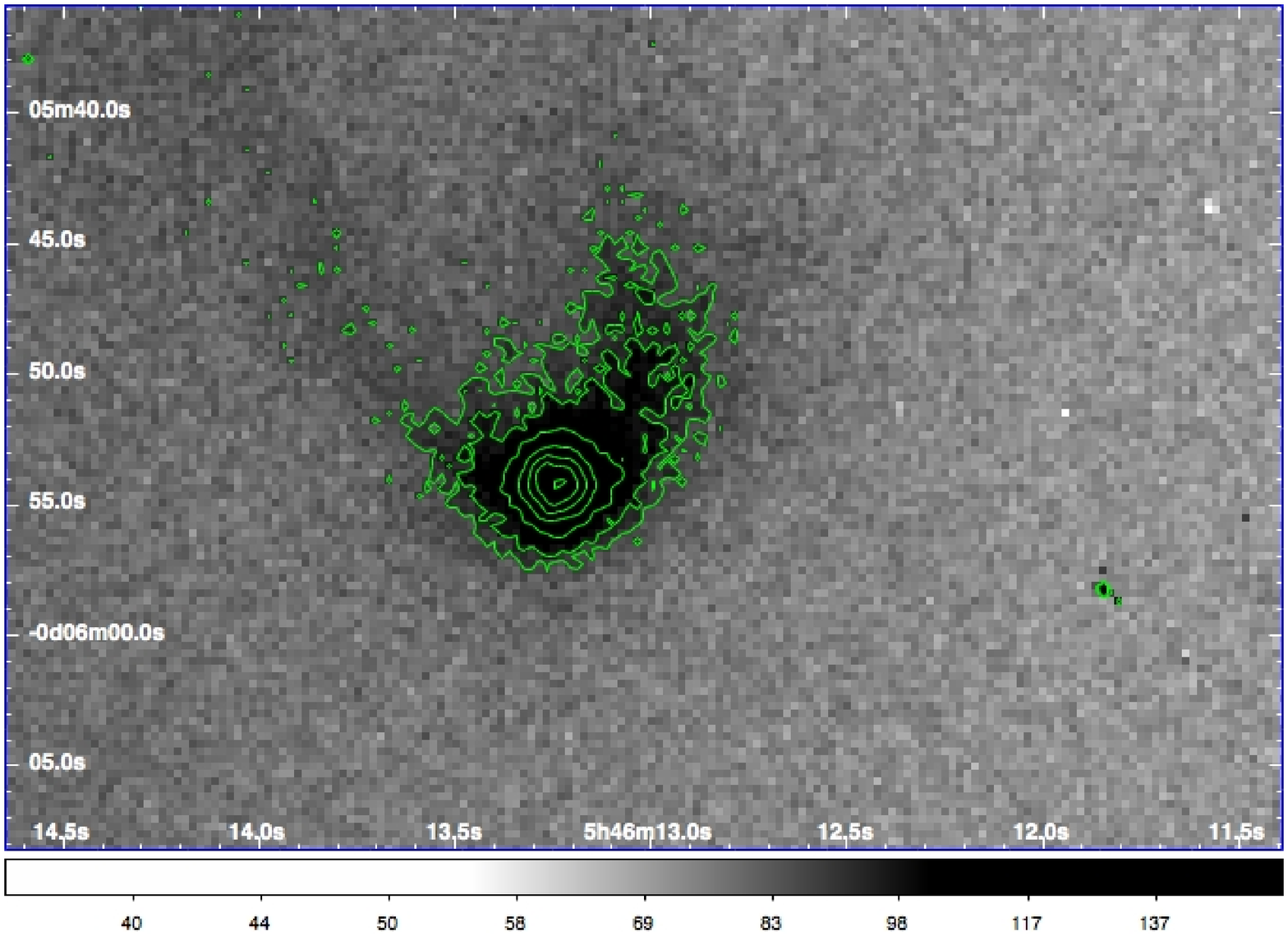}
\includegraphics[scale=0.30, angle=270]{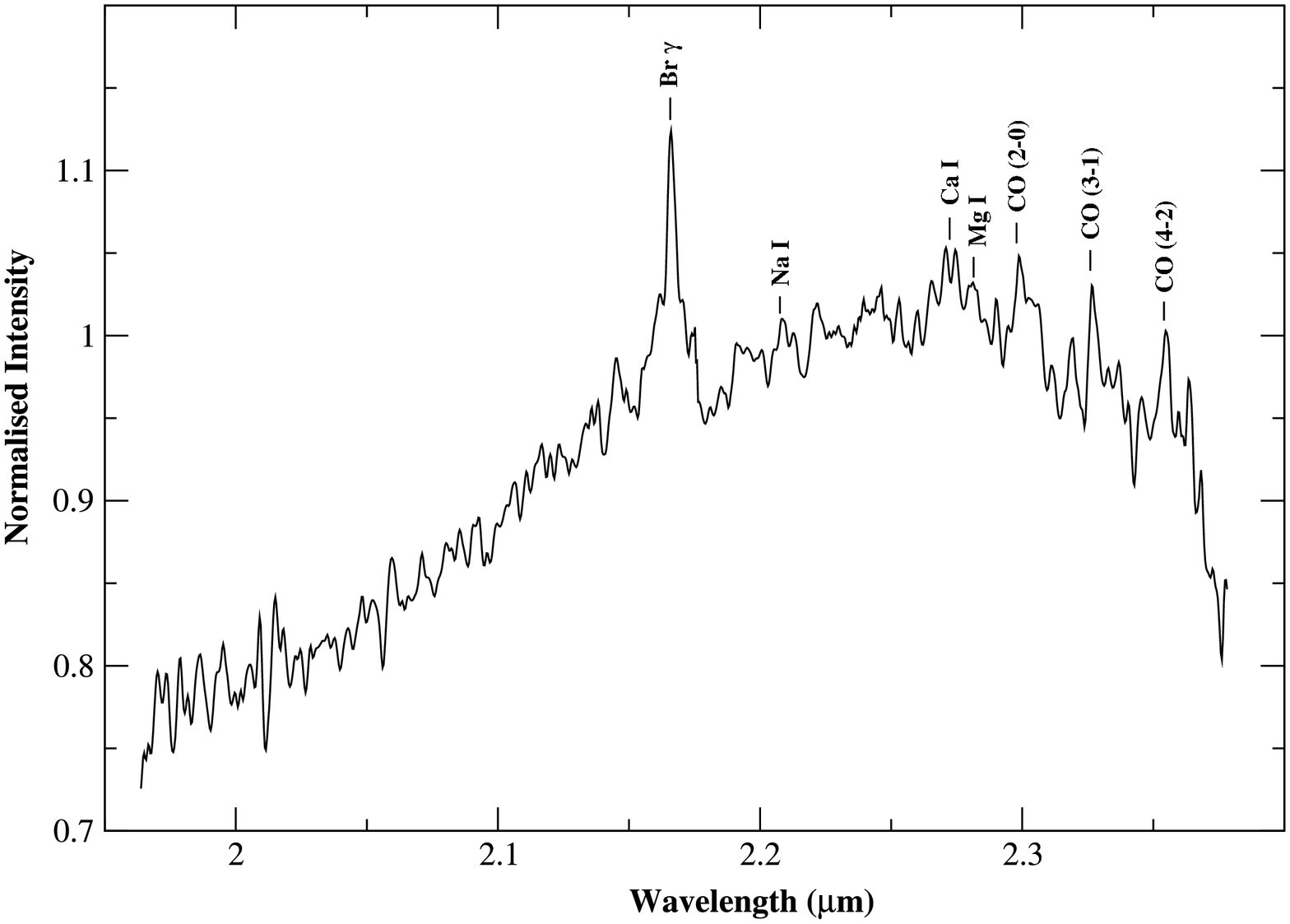}
\end{center}
\caption{Top panel: Chandra/ACIS-I image of the region surrounding McNeil's Nebula (Sept 2008). For comparison purposes we overlay a R-band color coded contour map of the nebula obtained with VLT/FORS2 (Oct2008). A bright embedded X-ray source is spatially coincident with V1647 Ori at the apex of McNeil's nebula. Middle panel:  Compact nebula seen around the illuminating star of McNeil's Nebula in this J-band image obtained with SOFI at the 3.6m telescope in La Silla on September 2008. Bottom panel: K-band spectrum of V1647 Ori, which displays a red continuum with strong CO band head emission, and the Br-$\gamma$ and Na I lines are in emission. The spectrum was taken with the 1.2m Mt Abu Observatory, India}
\end{figure}
\begin{figure}[]
\includegraphics[scale=0.4, angle=270]{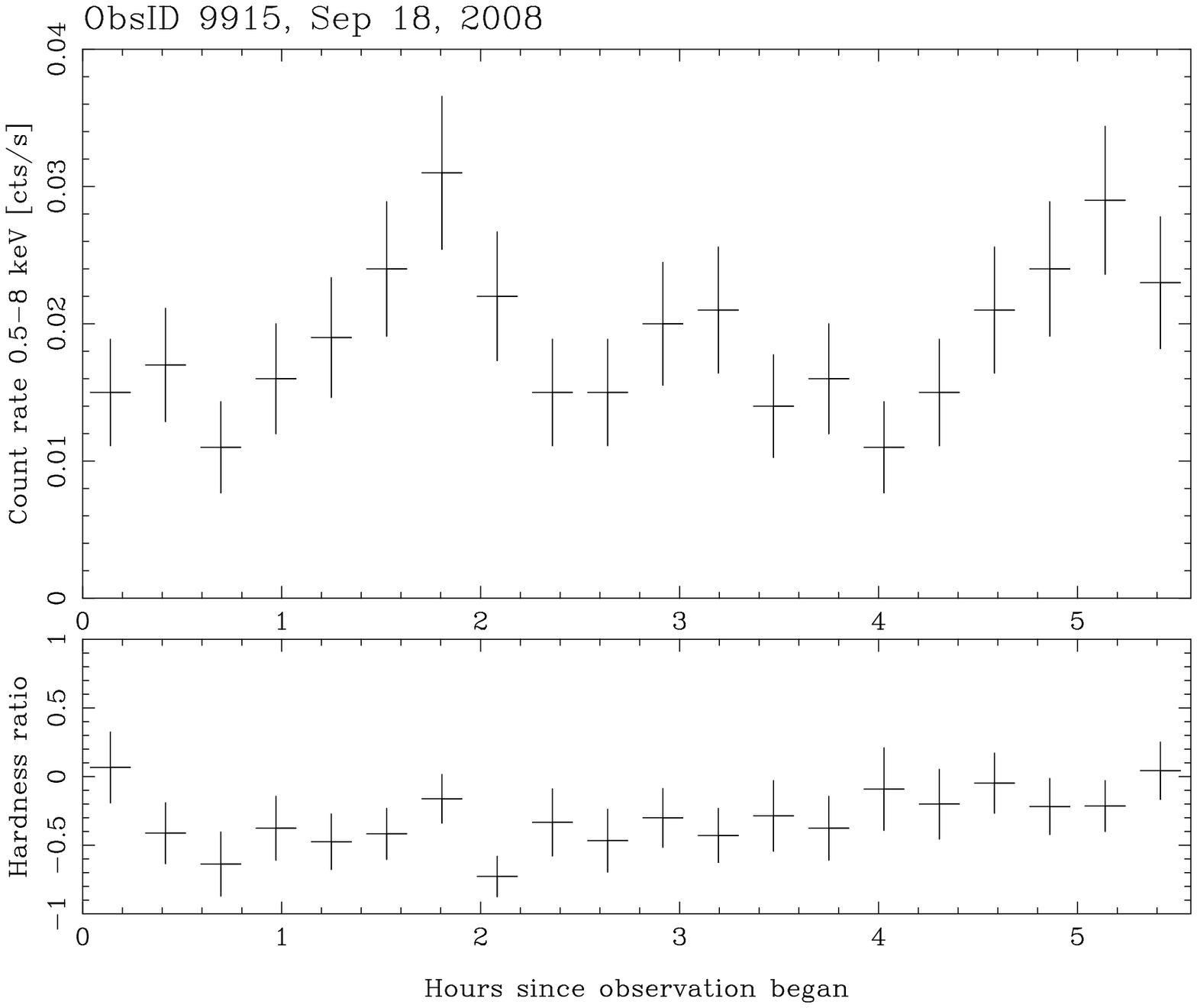}
\includegraphics[scale=0.4, angle=270]{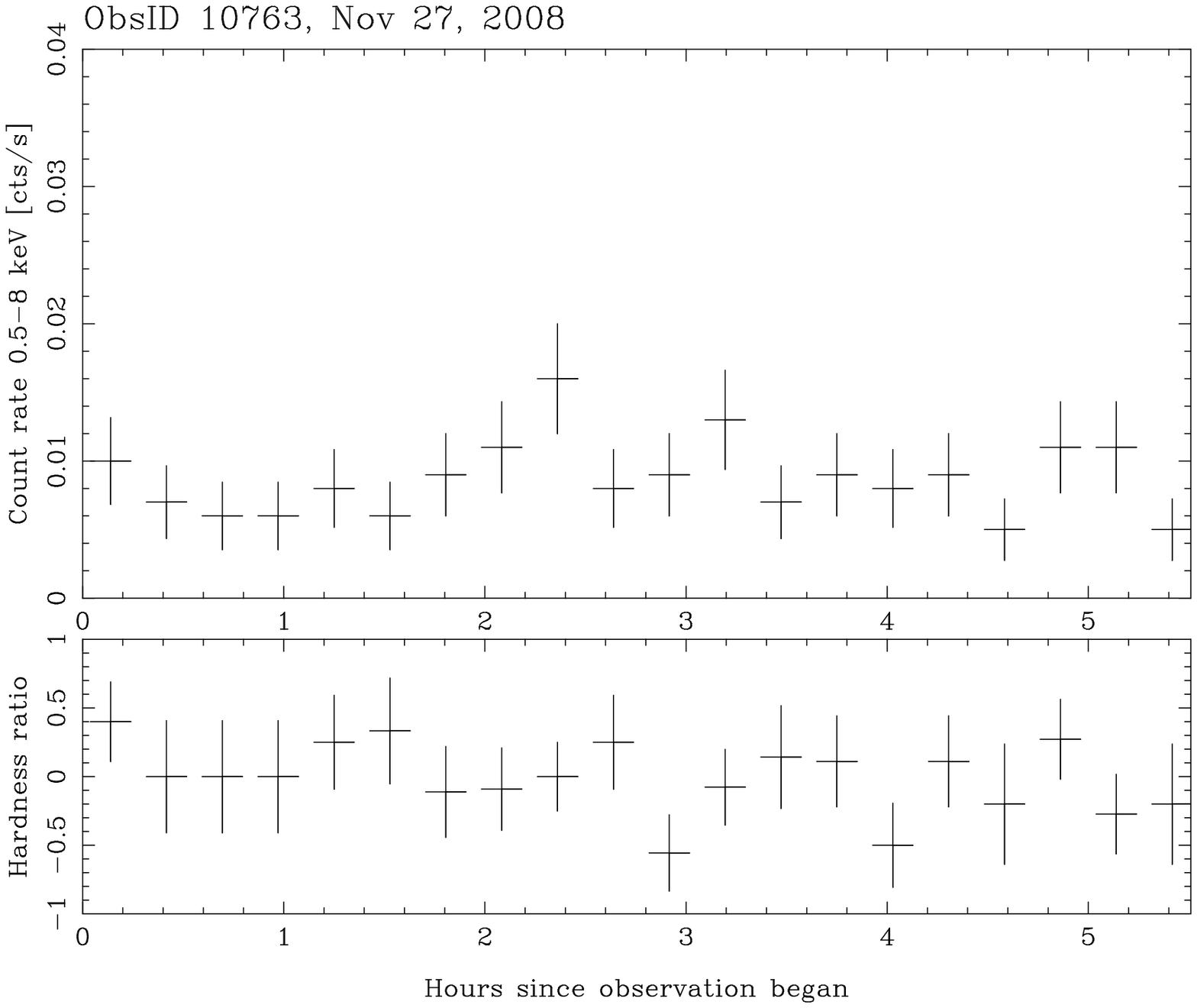}
\includegraphics[scale=0.4, angle=270]{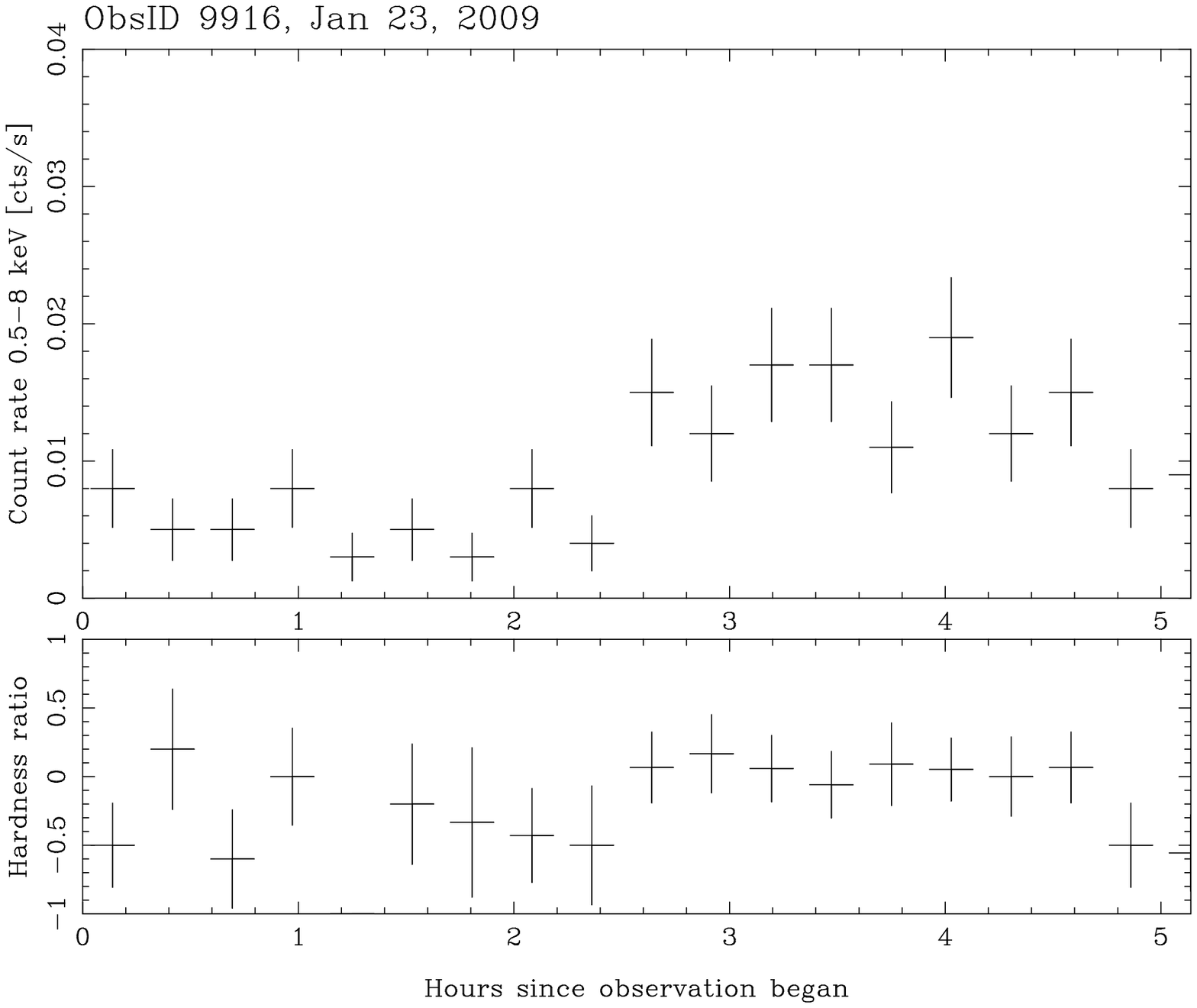}
\includegraphics[scale=0.4, angle=270]{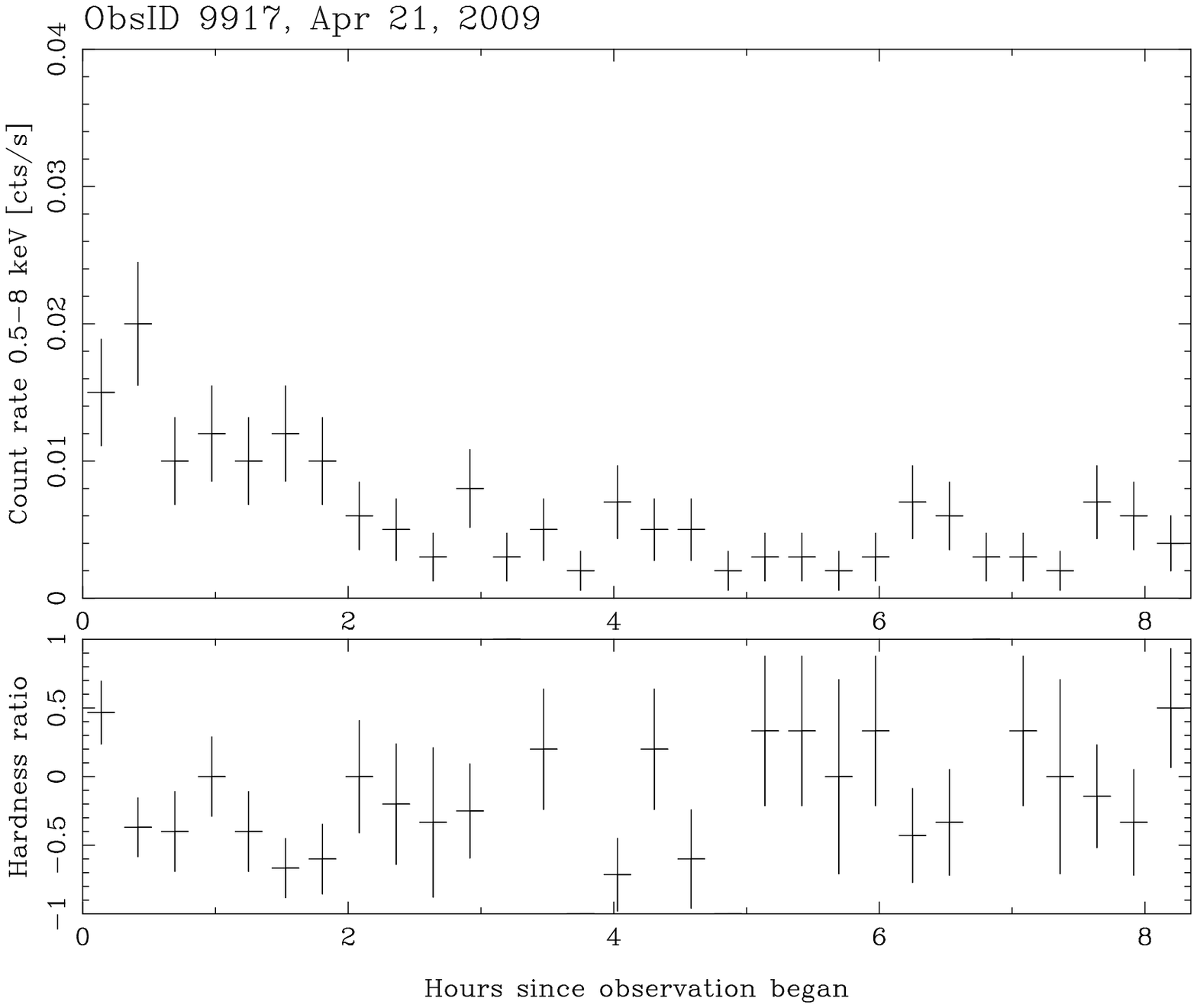}
\includegraphics[scale=0.065]{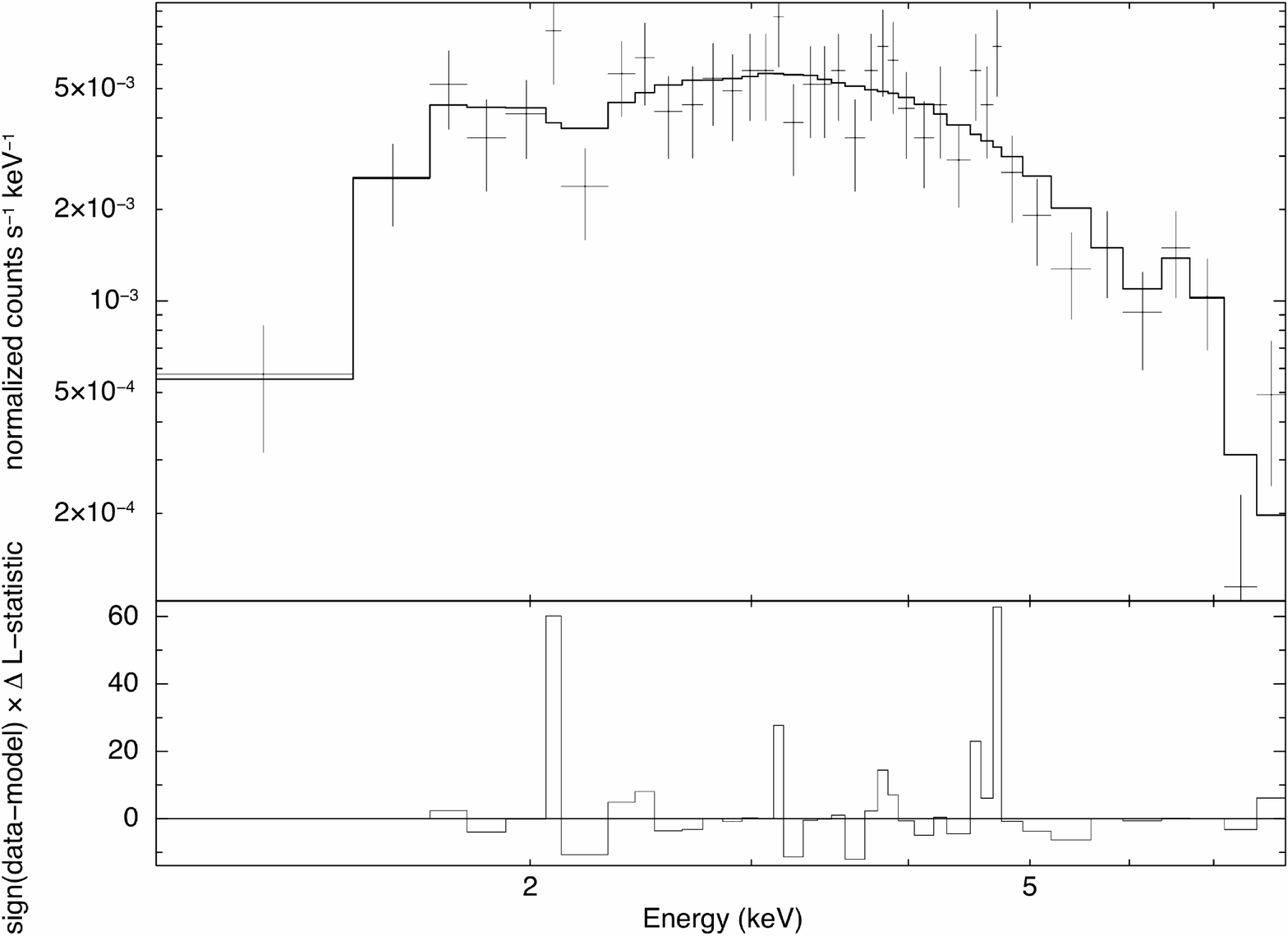}
\includegraphics[scale=0.065]{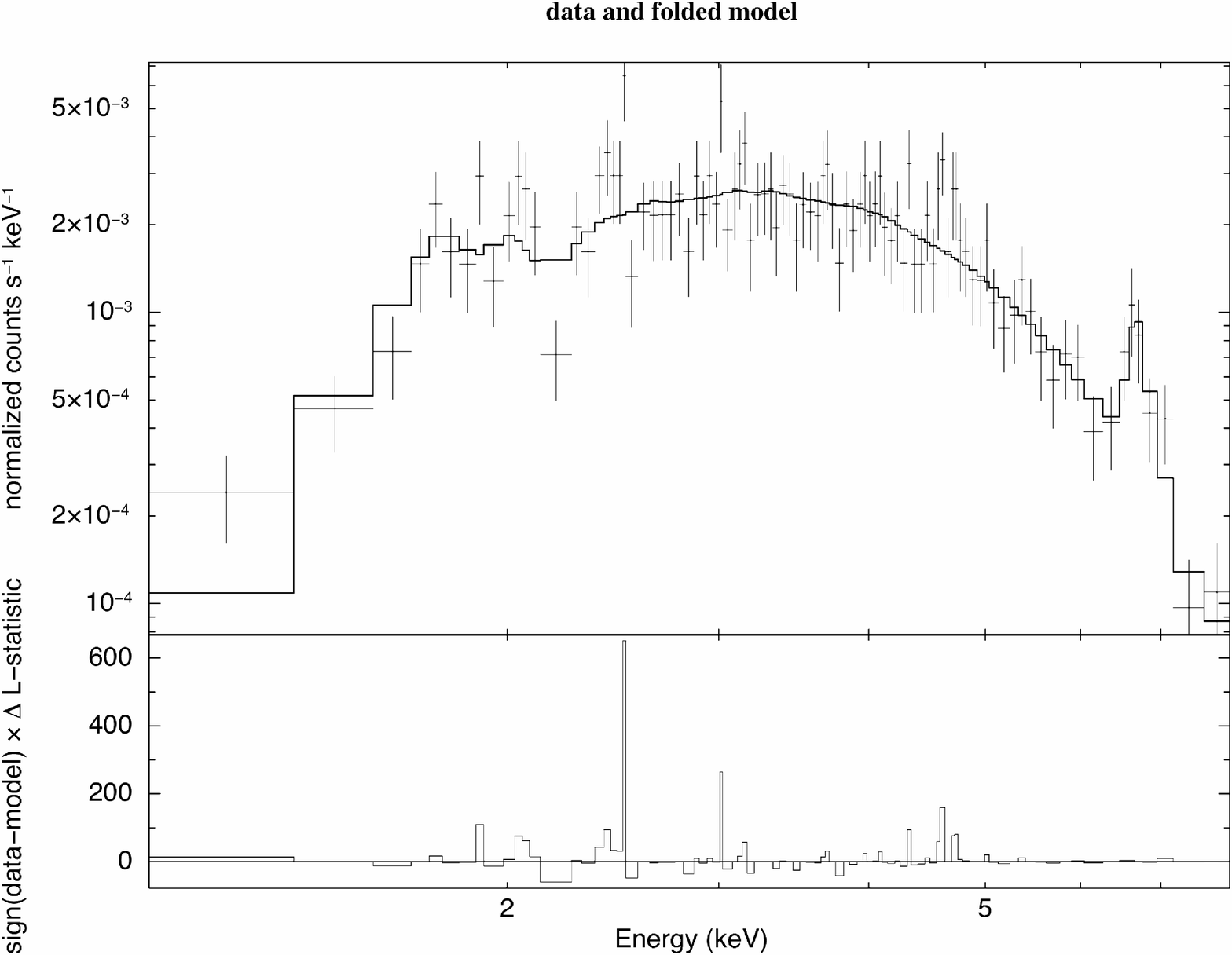}
\caption{Top four panels: Chandra/ACIS-I image background subtracted X-ray light curves (bin=1000ks) of V1647 Ori. The upper panels shows the ACIS-I X-ray light curve of V1647 Ori with one sigma error bars in the energy band from~0.5~to~8.0~keV. The bottom panels show the variation of the corresponding hardness ratio ( S=0.5-2.8keV and H=2.8-8keV). Bottom two panels: Chandra X-ray spectra and our best fit models of V1647 Ori. The upper panel shows the unbinned data15~counts in each spectral bin. The solid line show the best fit model with an absorbed single-temperature optically thin-thermal plasma emission with a Gaussian function at 6.4 keV for the iron fluorescent line. Bottom panels show the residuals from the best fit model. Left panel: Observation from September 2008 (outburst onset). Right panel: Merged of all observations available obtained between Sept 2008 and April 2009. It is clear the detection of the iron fluorescent line at 6.4keV. The equivalent with obtained was about 90eV similar to that reported by \citet{grosso05}.
}
\end{figure}

\section{Observations and Results}

\subsection{Optical and NIR Observations}
We studied the brightness and spectral evolution of the young eruptive star V1647 Ori during its recent outburst in the period 2008 August- 2010 August. We performed a photometric follow-up in the bands V, R, I, J, H, and K, as well as visible and near-IR spectroscopy. We also present new data for the 2004 February-2006 September outburst. We combined our data from the 2004 and 2008 outburst with those published by other authors. Our observations were carried out at: 1) NOT and WHT, in La Palma, Spain; 2) IAC80 and TCS, Tenerife, Spain; 3) 1m RCC, Konkoly Obs., Hungary; 4) 2.2m, Calar Alto, Spain; 5) 1.2m PRL, Mt Abu, India; 6) 2m RCC, NAO Rozhen, Bulgaria. 
For the optical images we performed a standard data reduction process (bias subtraction, correction for flatfield and cosmic rays) using IRAF. The magnitude was computed using differential aperture photometry centered at the stellar position. The sky has been computed within an annulus and subtracted from the aperture photometry. For the spectroscopic data, the spectra were reduced using standard iraf routines. The one-dimensional spectra were extracted from the bias-subtracted and flatfield corrected images using the optimal extraction method. Wavelength calibration of the spectra was done using the arc lamp sources. Subsequently, the spectra were corrected for instrumental response using spectrophotometric standards.

Fig.~5 clearly shows that the star is surrounded by a compact reflection nebulosity and that, in particular, this nebula shows a curved tail characteristic of many stars undergoing high-accretion events \citep[e.g.][]{herbig77}.

We also present Spitzer Space Telescope observations of V1647 Ori. The outbursting source was observed several times between the optical peak of the outburst in early March 2004 and March 2005. We also present observations from Novemeber 2009 related to the 2008 outburst. In both cases the source is easily detected in all Spitzer imaging bands from 3.6 to 8.0 microns. The fluxes at all wavelengths are roughly a factor of 15 brighter than pre-outburst levels. The linearization, flatfielding, and dark subtraction of the data were performed by using the IRAC pipeline at the Spitzer Science Center. 
\subsection{X-ray Observations}
The Chandra X-ray satellite observed V1647 Ori on 2008 five times between September 2008 and April 2009 (ObIDs 8585, 9915, 9916, 9917, and 10763). All the observations employed the ACIS-I detector in its standard instrument configuration. Pipeline-processed (CXC, ver. 4.3.1) photon event lists were reduced using the CIAO software package, version 4.2. V1647 Ori showed a level of X-ray emission two orders of magnitude larger than that of the quiescent state. Such enhanced X-ray variability was also seen in XMM-Newton observations in 2004 and 2005 during the 2004 outburst, but has rarely been observed for other young stellar object. The spectrum clearly displays emission from Helium-like iron, which is a signature of hot plasma (kT$\sim$5 keV). The best-fit plasma temperature ($\sim$4.65 keV), elemental abundance ($\sim$0.50 solar), hydrogen column density ($\sim$4x10$^{22}$ erg/cm2/s), and the average observed flux ($\sim$7x10$^{-13}$ erg/cm2/s between 0.5-8.0 keV) shown in Fig~6 are similar to those inferred for V1647 Ori by \citet{grosso05} and \citet{hamaguchi10} during the 2004 and 2008 outburst respectively. The spectrum also shows a fluorescent iron K$\alpha$ line with a equivalent width (EW) of $\sim$90eV, similar to that reported by \citet{grosso05}, which suggest that the structure of the circumstellar was gas very close to the stellar core that absorbs and re-emits X-ray emission from the central object. \citet{hamaguchi10}, using Suzaku X-ray satellite  obtained a EW of $\sim$600 eV in a single exposure around the time of the outburst onset. It suggests that a part of the incident X-ray emission that irradiates the circumstellar material and/or the stellar surface is hidden from our line of sight. 
\acknowledgements D.G.A. acknowledge support from the Spanish MICINN through grant AYA2008-02038 and the Ramón y Cajal Program ref. RYC-2005-000549. J. J. D. and V. K. were supported by CXC NASA contract NAS8-39073. This research has made use of the NASA's High Energy Astrophysics Science Archive Research Center (HEASARC) and observations made with ESO Telescopes at the La Silla and Paranal Observatories.
\bibliography{garciaalvarez_d_v2}

\begin{thebibliography}{}
\expandafter\ifx\csname natexlab\endcsname\relax\def\natexlab#1{#1}\fi
\expandafter\ifx\csname url\endcsname\relax
  \def\url#1{\texttt{#1}}\fi
\expandafter\ifx\csname urlprefix\endcsname\relax\def\urlprefix{URL }\fi
\providecommand{\eprint}[2][]{\url{#2}}

\bibitem[{{Acosta-Pulido} et~al.(2007){Acosta-Pulido}, {Kun},
  {{\'A}brah{\'a}m}, {K{\'o}sp{\'a}l}, {Csizmadia}, {Kiss}, {Mo{\'o}r},
  {Szabados}, {Benk{\H o}}, {Barrena Delgado}, {Charcos-Llorens}, {Eredics},
  {Kiss}, {Manchado}, {R{\'a}cz}, {Ramos Almeida}, {Sz{\'e}kely}, \&
  {Vidal-N{\'u}{\~n}ez}}]{acosta-pulido07}
{Acosta-Pulido}, J.~A., {Kun}, M., {{\'A}brah{\'a}m}, P., {K{\'o}sp{\'a}l},
  {\'A}., {Csizmadia}, S., {Kiss}, L.~L., {Mo{\'o}r}, A., {Szabados}, L.,
  {Benk{\H o}}, J.~M., {Barrena Delgado}, R., {Charcos-Llorens}, M., {Eredics},
  M., {Kiss}, Z.~T., {Manchado}, A., {R{\'a}cz}, M., {Ramos Almeida}, C.,
  {Sz{\'e}kely}, P., \& {Vidal-N{\'u}{\~n}ez}, M.~J. 2007, \aj, 133, 2020

\bibitem[{{Aspin} et~al.(2008){Aspin}, {Beck}, \& {Reipurth}}]{aspin08}
{Aspin}, C., {Beck}, T.~L., \& {Reipurth}, B. 2008, \aj, 135, 423

\bibitem[{{Brice{\~n}o} et~al.(2004){Brice{\~n}o}, {Vivas}, {Hern{\'a}ndez},
  {Calvet}, {Hartmann}, {Megeath}, {Berlind}, {Calkins}, \&
  {Hoyer}}]{briceno04}
{Brice{\~n}o}, C., {Vivas}, A.~K., {Hern{\'a}ndez}, J., {Calvet}, N.,
  {Hartmann}, L., {Megeath}, T., {Berlind}, P., {Calkins}, M., \& {Hoyer}, S.
  2004, \apjl, 606, L123

\bibitem[{{Chochol} et~al.(2006){Chochol}, {Errico}, {Magr{\`i}}, {Pribulla},
  \& {Vittone}}]{chochol06}
{Chochol}, D., {Errico}, L., {Magr{\`i}}, M., {Pribulla}, T., \& {Vittone},
  A.~A. 2006, Contributions of the Astronomical Observatory Skalnate Pleso, 36,
  149

\bibitem[{{Cohen} et~al.(1981){Cohen}, {Kuhi}, {Spinrad}, \&
  {Harlan}}]{cohen81}
{Cohen}, M., {Kuhi}, L.~V., {Spinrad}, H., \& {Harlan}, E.~A. 1981, \apj, 245,
  920

\bibitem[{{Fedele} et~al.(2007){Fedele}, {van den Ancker}, {Petr-Gotzens}, \&
  {Rafanelli}}]{fedele07}
{Fedele}, D., {van den Ancker}, M.~E., {Petr-Gotzens}, M.~G., \& {Rafanelli},
  P. 2007, \aap, 472, 207

\bibitem[{{Grosso} et~al.(2005){Grosso}, {Kastner}, {Ozawa}, {Richmond},
  {Simon}, {Weintraub}, {Hamaguchi}, \& {Frank}}]{grosso05}
{Grosso}, N., {Kastner}, J.~H., {Ozawa}, H., {Richmond}, M., {Simon}, T.,
  {Weintraub}, D.~A., {Hamaguchi}, K., \& {Frank}, A. 2005, \aap, 438, 159

\bibitem[{{Hamaguchi} et~al.(2010){Hamaguchi}, {Grosso}, {Kastner},
  {Weintraub}, \& {Richmond}}]{hamaguchi10}
{Hamaguchi}, K., {Grosso}, N., {Kastner}, J.~H., {Weintraub}, D.~A., \&
  {Richmond}, M. 2010, \apjl, 714, L16

\bibitem[{{Herbig}(1977)}]{herbig77}
{Herbig}, G.~H. 1977, \apj, 217, 693

\bibitem[{{Kaurav} et~al.(2010){Kaurav}, {Ojha}, {Ninan}, {Bhatt}, {Sahu}, \&
  {Ghosh}}]{kaurav10}
{Kaurav}, S.~S., {Ojha}, D.~K., {Ninan}, J.~P., {Bhatt}, B.~C., {Sahu}, D.~K.,
  \& {Ghosh}, A., S.~K.and~{Tej} 2010, Contributions of the Astronomical
  Observatory Skalnate Pleso, 1, 203

\bibitem[{{Kospal} et~al.(2005){Kospal}, {Abraham}, {Csizmadia}, {Eredics},
  {Kun}, \& {Racz}}]{kospal05}
{Kospal}, A., {Abraham}, A., {Csizmadia}, S., {Eredics}, M., {Kun}, M., \&
  {Racz}, M. 2005, Information Bulletin on Variable Stars, 5661, 1

\bibitem[{{Kun}(2008)}]{kun08}
{Kun}, M. 2008, Information Bulletin on Variable Stars, 5850, 1

\bibitem[{{McGehee} et~al.(2004){McGehee}, {Smith}, {Henden}, {Richmond},
  {Knapp}, {Finkbeiner}, {Ivezi{\'c}}, \& {Brinkmann}}]{mcgehee04}
{McGehee}, P.~M., {Smith}, J.~A., {Henden}, A.~A., {Richmond}, M.~W., {Knapp},
  G.~R., {Finkbeiner}, D.~P., {Ivezi{\'c}}, {\v Z}., \& {Brinkmann}, J. 2004,
  \apj, 616, 1058

\bibitem[{{McNeil} et~al.(2004){McNeil}, {Reipurth}, \& {Meech}}]{mcneil04}
{McNeil}, J.~W., {Reipurth}, B., \& {Meech}, K. 2004, IAU Circ., 8284, 1

\bibitem[{{Meyer} et~al.(1997){Meyer}, {Calvet}, \& {Hillenbrand}}]{meyer97}
{Meyer}, M.~R., {Calvet}, N., \& {Hillenbrand}, L.~A. 1997, \aj, 114, 288

\bibitem[{{Muzerolle} et~al.(2005){Muzerolle}, {Megeath}, {Flaherty}, {Gordon},
  {Rieke}, {Young}, \& {Lada}}]{muzerolle05}
{Muzerolle}, J., {Megeath}, S.~T., {Flaherty}, K.~M., {Gordon}, K.~D., {Rieke},
  G.~H., {Young}, E.~T., \& {Lada}, C.~J. 2005, \apjl, 620, L107

\bibitem[{{Ojha} et~al.(2008){Ojha}, {Ghosh}, {Kaurav}, {Bhatt}, {Sahu}, \&
  {Tej}}]{ojha08}
{Ojha}, D.~K., {Ghosh}, S.~K., {Kaurav}, S.~S., {Bhatt}, B.~C., {Sahu}, D.~K.,
  \& {Tej}, A. 2008, IAU Circ., 9006, 1

\bibitem[{{Ojha} et~al.(2006){Ojha}, {Ghosh}, {Tej}, {Verma}, {Vig}, {Anupama},
  {Sahu}, {Parihar}, {Bhatt}, {Prabhu}, {Maheswar}, {Bhatt}, {Anandarao}, \&
  {Venkataraman}}]{ojha06}
{Ojha}, D.~K., {Ghosh}, S.~K., {Tej}, A., {Verma}, R.~P., {Vig}, S., {Anupama},
  G.~C., {Sahu}, D.~K., {Parihar}, P., {Bhatt}, B.~C., {Prabhu}, T.~P.,
  {Maheswar}, G., {Bhatt}, H.~C., {Anandarao}, B.~G., \& {Venkataraman}, V.
  2006, \mnras, 368, 825

\bibitem[{{Reipurth} \& {Aspin}(2004)}]{reipurth04}
{Reipurth}, B., \& {Aspin}, C. 2004, \apjl, 606, L119

\bibitem[{{Rieke} \& {Lebofsky}(1985)}]{rieke85}
{Rieke}, G.~H., \& {Lebofsky}, M.~J. 1985, \apj, 288, 618

\bibitem[{{Semkov}(2006)}]{semkov06}
{Semkov}, E.~H. 2006, Information Bulletin on Variable Stars, 5683, 1

\end{thebibliography}

\end{document}